\documentclass[onecolumn,letterpaper]{aa}
\usepackage{graphicx}
\usepackage[sort&compress]{natbib}
\usepackage{txfonts}
\begin{document}
   \title{A new design for a very low frequency space borne radio interferometer}


   \author{Divya Oberoi \inst{1} \and Jean-Louis Pin\c{c}on \inst{2} }

   \offprints{Divya Oberoi}

   \institute{Haystack Observatory, 
	      Massachusetts Institute of Technology, 
              Off Route 40, Westford, MA, 01886, USA\\
              \email{doberoi@haystack.mit.edu} 
              \and
	      Laboratoire de Physique et Chimie de l'Environnement,
              Centre Nationale de la Recherche Scientifique,
              3A, Avenue de la Recherche Scientifique,
              45071, Orleans Cedex 2, France \\
              \email{jlpincon@cnrs-orleans.fr}
             }

   \date{Received August 1, 2002; accepted September 1, 2002}

   \abstract{The non transparency and severe propagation effects of the
terrestrial ionosphere make it impossible for Earth based instruments to study
the universe at low radio frequencies.
An exploration of the low frequency radio window with the resolution and
sensitivity essential to meet the scientific objectives will necessarily require
a dedicated satellite based interferometer operating at these frequencies.
Such missions have been proposed in the literature for about past fifteen
years.
Today, the steady and impressive advances in technology and computing resources
have brought us on the brink of a quantum jump in the performance and capabilities
of such missions, increasing their scientific desirability many-fold.
This paper presents the concept design which emerged from a study to
investigate the feasibility of a low frequency satellite based interferometer
operating in the frequency range $0.1$ -- $40\ MHz$ titled
{\em PrepAration for Radio Interferometry in Space (PARIS)}.
The salient features of the design are: an on-board correlator to reduce the 
data volumes to be transmitted to the Earth by about two orders of magnitude; 
use of three orthogonal dipoles in place of two to achieve better 
polarisation characteristics; direct digitisation of the entire radio 
frequency band of interest which eliminates the need for a local oscillator 
system; an overlap in the observing frequency range with upcoming ground
based instruments for aid in imaging and calibration and an all-sky imaging
capability.
The most constraining bottle neck for the present design is the large
intra-constellation telemetry requirement.
It is expected that technological solutions to meet this requirement will
be found in near future as other formation flying missions which share this
requirement emerge.

   \keywords{
             Telescopes --
         Instrumentation: interferometers --
             Techniques: interferometric --
         Methods: observational
               }
   }
   \titlerunning{A new design for a VLF space borne radio interferometer}
   \authorrunning{D. Oberoi, J.L. Pin\c{c}on}
   \maketitle


\section{Introduction}
\label{sec:intro}

High resolution and high sensitivity low frequency radio astronomy
has been a standing challenge till the present time. On the Earth,
the non-transparency of the ionosphere below a few MHz and its
severe propagation effects below few tens of MHz have prevented
detailed studies of the radio universe at these frequencies. The
60s and 70s saw some attempts to investigate the
electromagnetic spectrum below 30 $MHz$ using space borne radio
astronomy experiments on-board individual satellites \citep[][and
references therein] {1971IAUS...41..401A}. The last explorations
in this series were conducted by the two Radio Astronomy Explorer
missions (RAE) in the late 1960s and early 1970s
\citep{1973PSSc...21..443, 1975A&A....40..365A}. These missions
were dedicated radio astronomy missions and covered the frequency
ranges $0.2$ -- $9.2\ MHz$ and $0.025$ -- $13.1\ MHz$, respectively
and had practically identical instrumentation.
RAE-1 was placed in a $6000\ km$, circular, $59^{\circ}$
inclination, retrograde orbit around the Earth and RAE-2 in
circular Lunar orbit at $1100\ km$ and $59^{\circ}$ inclination.
These satellites provided resolutions of about $40^{\circ}\! \times
\!60^{\circ}$ at $\sim\!9\ MHz$ and $220^{\circ}\! \times\! 160^{\circ}$
at $\sim\!1\ MHz$. Even today, it is not possible to achieve
significantly better resolution using individual spacecraft.

A space based interferometer will be essential to conduct a
detailed study of the radio universe below few tens of $MHz$. In
this paper, we refer to the frequency range below $30\ MHz$ as
the Very Low Frequency (VLF) range. The subject has been discussed 
in the literature for about fifteen years now.
\citet{1988A&A...195..372W} proposed a four satellite mission with
$85\ m$ crossed travelling wave {\em V}-antennas, similar to the
ones used on the RAE satellites. Four discreet bands of $50\ kHz$
each were to be covered in the range $1$ -- $30\ MHz$.
\citet{1997RaSc...32..combined} suggested a strategic plan for a
space based low frequency array progressing from spectral analysis
on-board a single spacecraft to a two element interferometer in
Earth orbit, continuing to an interferometer array in Earth or
Lunar orbit, culminating in Lunar nearside and far-side arrays.
They considered it necessary to have high gain antennas and
suggested the use of spherical inflatable arrays with a large
number of active elements as interferometer elements for the space
array. \citet{2000GMS...119..339J} were the first to consider
short dipoles to be suitable elements for space interferometer.
They proposed a 16 element array on a distant retrograde orbit,
covering $0.3$ -- $30\ MHz$ with up to $125\ kHz$ bandwidths of
observation. These studies focused on achieving the best possible
performance using the technology available to them.

The enormous advances in technology and the vast increase in the
computing resources available have brought it within reach to
tailor the solutions to the specific needs of VLF radio astronomy
and made it possible to plan the next generation of radio
interferometers. This is amply reflected in the many new projects
which are being actively pursued, for instance the Low Frequency
Interferometer (LOFAR), Square Kilometer Array (SKA), Alan
Telescope Array (ATA) and the Frequency Agile Solar Telescope
(FASR). This work is an attempt to present the next generation of
design for space based VLF interferometers. 
An important
guiding principle for the study was to come up with a design
tailored to the specific needs to VLF interferometry. A design
motivated by how one would like an ideal instrument to be, and not
one which can necessarily be realised in immediate future. The
limitations imposed by the currently available technology were
hence not regarded as hard constraints. We have, however,
exercised caution, in our endeavours to think beyond the current
technological limitations, to not wander off far away from the
realm of the feasible. The design presented is rather aggressive
in the aspects of technology which are progressing most rapidly
and practically respects the existing constraints from
technologies which are progressing at a slow pace. Judging from
present trends, it is expected to become feasible in 
near future. The study, titled {\em PrepAration for Radio 
Interferometry in Space (PARIS),} was a collaborative effort between 
the Laboratoire de Physique et Chimie de l'Environnement, the
Laboratoire d'Etudes Spatiales et d'Instrumentation en
Astrophysique from the Paris-Meudon Observatory and the Nan\c{c}ay
radio astronomy station.

The next section briefly summarises the scientific motivation
behind a VLF space interferometer. The third section details the
aspects of VLF sky and interferometry technique which were
regarded as the major design drivers for this study. The concept
design is presented in the fourth section and the fifth highlights
some key aspects of the data analysis strategy. Calibration and
formation flying issues are discussed in the sixth section and the
seventh section presents the conclusions.


\section{Scientific objectives}
\label{sec:sci-objectives}

\begin{figure}
\begin{flushleft}
\includegraphics[scale=0.645, angle=0]{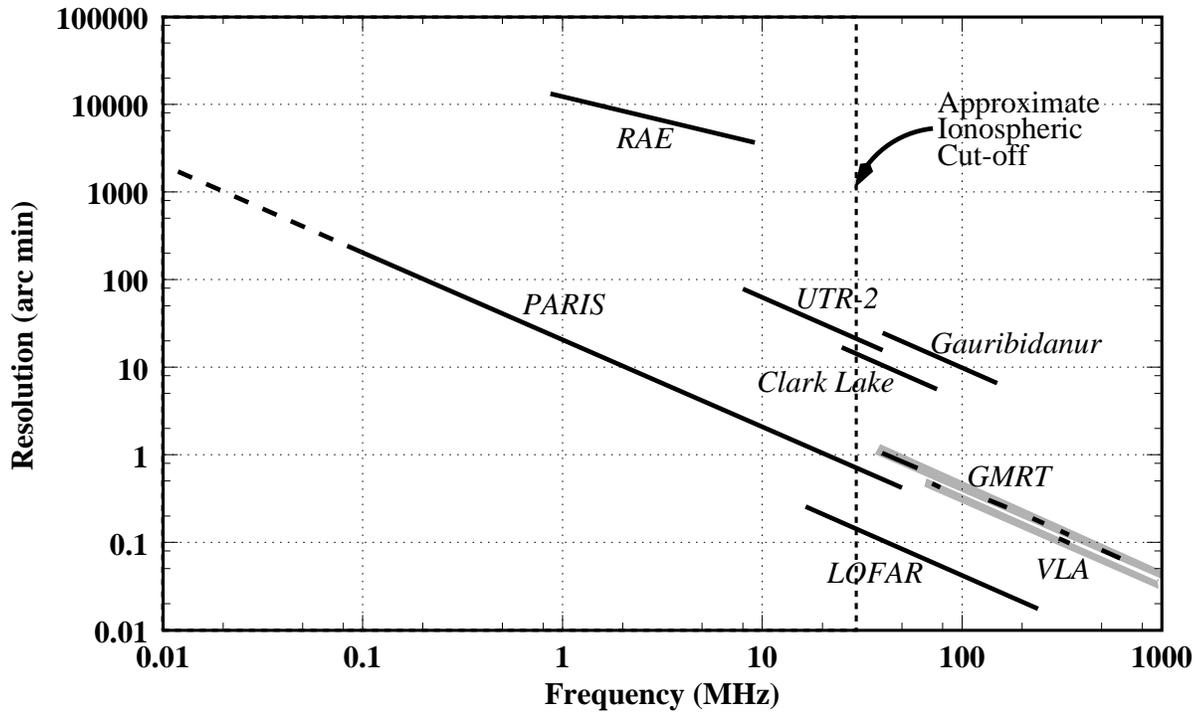}
\caption{{\bf The frequency coverage and resolution of various low
frequency radio telescopes} -- The figure clearly shows that the
PARIS proposal will provide unprecedented resolution and
frequency coverage in an unexplored region of the spectrum and
hence has considerable discovery potential. Approximate
Ionospheric cut-off frequency is also shown, below which it
becomes increasingly difficult to use ground based telescopes. }
\label{fig:LFinst}
\end{flushleft}
\end{figure}

Far from showing a simple extension of the more energetic
phenomena seen at higher frequencies, studies at very low radio
frequencies promise new insights in astrophysics and solar
physics. The main objective of PARIS is to produce the first ever
sensitive high resolution radio images of the entire sky in the
frequency range from $\sim\!300\ kHz$ to $\sim\!30\ MHz$.
Only a space borne radio interferometer, free from the corruptions due to
the ionosphere and away from the terrestrial radio frequency
interference, can deliver the sensitivity and the dynamic range
necessary for useful astronomy in this frequency range.
Fig.~\ref{fig:LFinst} clearly shows that PARIS will open up a new window 
of opportunity in the last unexplored part of the electromagnetic spectrum.
The topics discussed in this section highlight only some of the
major scientific objectives and potentials of the mission. 
For a more detailed review the reader is referred to the following 
dedicated publications: Radio Astronomy from Space by 
\cite{1987Radio_aston-space}; Low Frequency Astrophysics from Space by 
\cite{1990Low-freq-ap-space}; Radio Astronomy at Long Wavelengths by 
\cite{2000Radio-astro-at-long-wave}.

\subsection{Very low frequency astrophysics}
\label{subsec:astro-objectives} 
The astrophysics issues addressed by a VLF space borne radio interferometer 
include studies of discreet extra-galactic, galactic and solar system objects 
to studies, the diffuse emission of galactic origin and those of the intervening 
Interstellar Medium (ISM).
The most prominent among these are:
\begin{enumerate}
\item {\em The determination of very low frequency spectra} -- The measurement 
of the low frequency radio spectral behaviour is important for improving our 
understanding of the prevailing physical conditions and emission and absorption 
process at work \citep{2000ralw.conf..237E}.
Many physical processes involved in emission and absorption of radiation are 
observable only at very low radio frequencies. 
\item {\em Evolution of Galaxies} -- A low frequency space interferometer will 
be particularly sensitive to the radio emission from low energy electrons which 
have long radiative lifetimes.  
These {\em old} electrons form an excellent probe for studying the history of radio 
sources and will allow one to detect the {\em fossil} radio plasma whose signatures
at higher frequencies are much fainter 
\citep{2001A&A...366...26E,1997ApJ...487L.135R}.
These electrons allow one to detect the early epochs of activity of radio
galaxies, trace their evolution in time and will help constraint the models of 
galactic evolution and of super-massive black holes in the galactic nuclei.
\item {\em The study of  the exchange of matter and energy between stars and the
interstellar medium} --
Within our Galaxy, the interferometer will observe the background Galactic 
synchrotron emission and the extragalactic objects overlaid with a complex 
absorption pattern generated by the diffuse ISM and the discreet objects like 
supernova remnants.
Absorption from dense clumps of ionized hydrogen and the warm diffuse component 
of the ionized ISM will be detectable 
\citep{2000ralw.conf..257D,1990lfas.work..121R}.
Multi-frequency all sky maps will allow one to do a tomographic study of the
distribution of ionized hydrogen. 
This will be a considerable new source of information for improving the existing
model for galactic distribution of free electrons like the one by 
\cite{2002CordesLazioNE2001}.
These maps will also help in determining the origin of the energy content of 
the ionized hydrogen. 
In addition to revealing the large-scale structure of the ionized ISM, it may 
also be possible to probe the small scale turbulence, by examining the propagation 
effects like an apparent increase in angular size and fluctuation in dispersion
measure measurements towards pulsars \citep{2000ralw.conf..105C}. 
This will help in our understanding of the injection and turbulent dissipation 
of energy within the ISM.
By imaging the synchrotron emission from the electrons in the shock regions such 
a mission will allow to test the theory that cosmic rays are accelerated in 
supernova remnants shocks \citep{2000ralw.conf..277D}. 
\item {\em The discovery of new phenomena} -- An interferometer in this frequency 
range will provide a two order of magnitude improvement in both sensitivity and 
resolution, when compared to the existing observations from individual spacecraft. 
As with any first exploration of a part of the electromagnetic spectrum,
this mission is justifiably expected to discover objects and phenomena not seen 
at higher frequencies.
\end{enumerate}

\subsection{Solar and Space physics}
\label{subsec:solar-objectives} 
PARIS will produce the first ever interferometric very low frequency radio images 
of the solar corona, solar transients like the coronal mass ejections (CMEs) and 
the shock-fronts associated with consequent interplanetary disturbances (IPDs).
The key issues in solar/space physics to be addressed by a space borne VLF radio 
interferometer are:
\begin{enumerate}
\item {\em Imaging of solar transients and studying their evolution} -- 
Solar transient phenomena such as solar flares, filament eruptions and CMEs are 
manifested by distinct types of non-thermal radio emission. 
The study of the spatial and temporal evolution of this emission is essential to 
a better understanding of  the Sun-Earth connection. 
The proposed mission will allow us to image and track these solar induced 
disturbances, particularly the CMEs, from the vicinity of the Sun all the way to 
1 AU.
This requires observing frequencies from tens of $kHz$ to tens of $MHz$ and hence
these measurements can only be made from space. 
The existing low frequency observations of the Sun are all made using individual 
spacecraft.
Their strength lies in their ability to provide wide bandwidth and high spectral 
and temporal resolution dynamic spectra, but this data cannot provide radio images.
A high resolution wide band imaging instrument with continuous spectral coverage
will provide information of the morphology and spatial distribution of emission 
and will be a fitting compliment to the dynamic spectrum studies 
\citep{2000ralw.conf..115D,2000ralw.conf..123G}.
Images of transient radio bursts are also of prime important for space weather 
forecasting applications. 
It is, perhaps, worth noting that the STEREO mission, scheduled for launch in late
2005, will track the centroids of radio emission and shed some light on the 
emission pattern of the emission but will not have true imaging capabilities.
\item {\em Mapping of large scale interplanetary magnetic field topology and 
interplanetary density structures} -- 
The mission will also provide the means for remote sensing of the coronal and
interplanetary density and magnetic field structures in the inner heliosphere. 
For instance the electron beams producing type III bursts follow magnetic field 
lines outward from their source in the corona. 
Imaging and tracking these bursts will reveal the topology of interplanetary 
magnetic field lines.
\item {\em Improving our understanding of emission mechanisms and particle acceleration} --
The extended dynamic range of such an instrument will make it possible to image 
both thermal and non-thermal sources simultaneously.
Very low frequency images of thermal emission will, for the first time, provide 
information on coronal holes, streamers and CME structures at heliocentric 
distances larger than 2 solar radii. 
This information will form a natural complement to and will extend the maps of 
these structures obtained by white light and X ray imaging.
Images of shock-fronts and bursts associated with solar energetic particle (SEP) 
events will help improve our understanding of particle acceleration sites and
mechanisms \citep{2003GeoRL..30lSEP6M}.
\end{enumerate}


\section{VLF interferometry}
\label{sec:vlf-interferometry}

Though all interferometers share the same mathematical
foundations, their practical implementations change considerably
with the frequency range of interest. Every few orders of
magnitude in wavelength, the problem of designing an
interferometer changes in character and essentially evolves into a
different problem with new and different aspects becoming the
design drivers. For instance, the detailed designs of an optical
interferometer and a high frequency radio interferometer do not
have much in common, even though they implement the same
functional blocks. The VLF range (say $0.1$ -- $30\ MHz$) differs
from the usual domain of radio astronomy (say $0.1$ -- $30\ GHz$)
by three orders of magnitude. It is reasonable to expect the
design considerations for a VLF interferometer to differ from
those for interferometers at much higher radio frequencies. Some
of these considerations stem from natural causes and others due to
from technological aspects. In this section we identify the key
aspects of VLF interferometry which were considered to be the
major design drivers.

\subsection{The VLF sky}
\label{subsec:vlf-sky} The most conspicuous feature of the VLF
sky is the enormously strong Galactic background radiation. Fig.
{\bf XX} shows both the brightness temperature and the specific
intensity in the frequency range of interest. The specific intensity
\begin{figure}
\begin{center}
\includegraphics[scale=0.5, angle=-90]{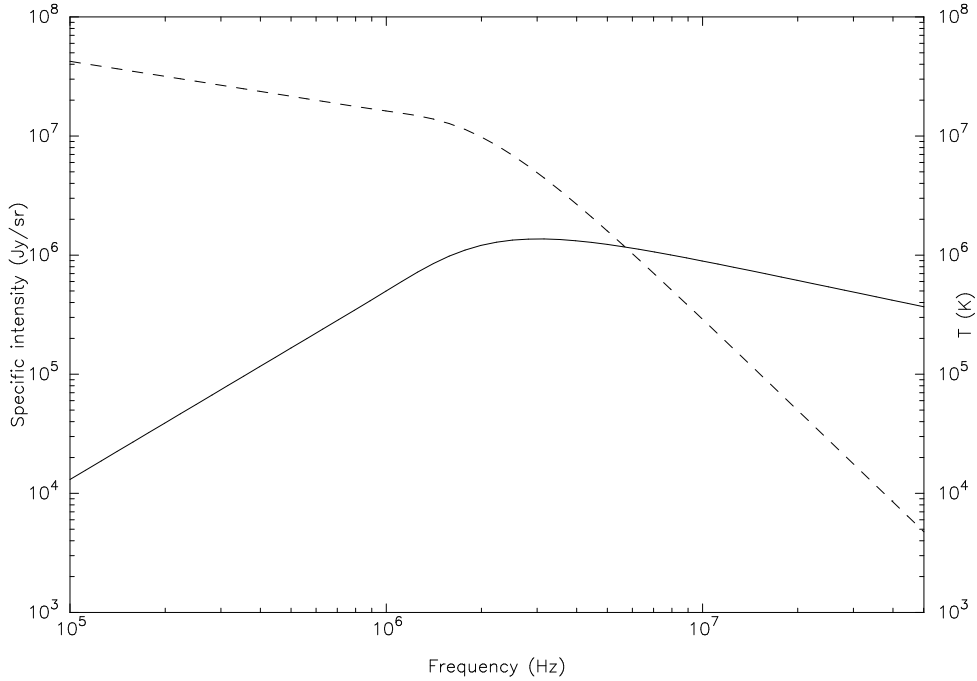}
\caption{{\bf The Galactic background radiation} -- The solid line 
shows the specific intensity of the Galactic background emission 
in $Jy/sr$ and the dashed line the corresponding temperature in 
degrees $K$.
}
\label{fig:gal_bg}
\end{center}
\end{figure}
of polar galactic background is $\sim\! 5\! \times\! 10^7\ Jy$ at $30\ MHz$, 
peaks at a value of $\sim\! 1.1\! \times\! 10^9 Jy$ close to $3\ MHz$ and slowly 
reduces to $\sim\! 8\!\times\! 10^6\ Jy$ by $0.3\ MHz$.
The brightness temperature is in excess of $\sim\! 10^4\ K$ by $30\ MHz$,
comes close to $\sim\! 10^7\ K$ at $3\ MHz$ and continues to rise more
slowly to $\sim\! 1.3\! \times\! 10^7$ at $0.3\ MHz$.
To put things in perspective, the polar galactic background brightness 
temperature at $408\ MHz$ is $\sim\!20\ K$ \citep{1982A&AS...47....1H} and 
reduces to $\sim\!42\ mK$ at $3.8\ GHz$ \citep{1998ApJ...505..473P}.

\subsection{Large field of view}
\label{subsec:largeFoV}
The directivity, or field of view (FoV), of a receptor for electromagnetic 
waves is characterised by $\lambda / d$, where $\lambda$ is the wavelength
of observation and $d$ the dimensions of the receptor. 
The Very Large Array (VLA), one of the most successful radio interferometers,
has antennas of 25 meter diameter and is most often used to observe
in a wavelength range from $0.224\ m$ to $0.0125\ m$ ($1.34$ -- $24\ GHz$).
Measured in units of $\lambda$, the diameter of the antennas ranges from 
$\sim\!112$ at the low frequency end to $\sim\!2000$ at the high end, leading 
to FoVs smaller than a hundredth to a thousandth of a radian across.
In the VLF band the $\lambda$ ranges from $1000\ m$ at $0.3\ MHz$ to $10\ m$
at $30 MHz$. 
The large wavelengths and the necessity to deploy the
structure in space preclude the possibility of limiting the field
of view of the receptors by using apertures many $\lambda$ in size, 
at least in the near future. 
The receptor size is expected to be smaller than the wavelength of operation for
practically the entire frequency range. The FoV of
individual receptors is hence expected to be very large.

The intense galactic background and the very large FoV imply that unlike at
high frequencies where the $T_{sys}$, the equivalent noise temperature 
corresponding to the sum of all the contributions to the signal received at the
output of an interferometer element, is dominated by $T_{rec}$, the noise 
contribution of the receiver electronics, at VLF frequencies, $T_{sys}$ will 
necessarily be very large and will be dominated by the contribution of the 
galactic background (Fig. \ref{fig:gal_bg}).

\subsection{Three dimensional sampling}
\label{subsec:3Dsampling}
The aim of interferometric imaging is to arrive at the {\em Brightness 
distribution} in the sky at an observing frequency $\nu$, $I_{\nu}(l,m)$, 
from the measured visibilities, $V_{\nu}(u,v,w)$.  
For a phase tracking interferometer with a small fractional bandwidth 
($\Delta \nu/\nu$), the two are related by the following expression
which gives the response to spatially incoherent radiation from
the far field 
\begin{equation}
V_{\nu}(u,v,w) = \int_{-\infty}^{\infty}\int_{-\infty}^{\infty} A(l,m)\ I_{\nu}(l,m)\ e^{-2\pi i\{ul+vm+w(\sqrt{1-l^2-m^2}-1)\}}\ \frac{dl\ dm}{\sqrt{1-l^2-m^2}}
\label{eq:visibity-brightness}
\end{equation}
where $u$, $v$ and $w$ are the orthogonal components of the baselines. 
They are measured in units of $\lambda$ and forming a right handed coordinate 
system such that $u$ and $v$ are measured in a plane perpendicular to the 
direction of the phase center, $u$ pointing to the local East and $v$ to the 
local North.
$l$ and $m$ are direction cosines measured with respect to the $u$-$v$-$w$ 
coordinate system and $A(l,m)$ is the antenna beam pattern.
If the third term in the exponential can be ignored, then equation 
\ref{eq:visibity-brightness} reduces to an exact 2D {\em Fourier} transform 
relationship \citep{1999sira.conf..383P}.
This can usually be achieved by limiting the FoV of the antenna primary beam to 
a narrow enough angular region, by building a large enough aperture. 
This is referred to as the {\em small FoV approximation} and most of the
existing interferometers operate in this regime.
Due to the large FoVs in the VLF regime, we will be well outside this
regime and hence the full 3D formalism will need to be employed for the 
inversion of visibility data.



For most synthesis imaging instrument in operation, due to the distribution
of the antennas on a near planar surface and small FOVs, it usually suffices 
to decompose the baseline vectors into the {\em u} and the {\em v} components.
For a space borne VLF interferometer with a 3D distribution of elements
and very large FoVs, it will be necessary to decompose the baselines along $u$, 
$v$ and $w$ axes.
Analogous to conventional ground based synthesis imaging, where the fidelity of
the final image depends upon the completeness of the sampling of
the $u$-$v$ plane, for a VLF space array it will depend upon the
completeness with which the $u$-$v$-$w$ volume is sampled.
The constellation configuration chosen for a VLF interferometer must
therefore try to achieve a good sampling of the 3D $u$-$v$-$w$ volume, as
opposed to the 2D $u$-$v$ plane as for most other aperture synthesis 
interferometers.

\subsection{Mapping the entire field of view}
\label{subsec:map-full-FoV}

The Eq. 1 in the previous section, truly holds only for a monochromatic 
interferometer.
All practical instruments measure visibilities over a finite bandwidth, 
$\Delta \nu$, centered at some frequency, $\nu_0$ and, the data within 
$\Delta \nu$ are treated as if they were at $\nu_0$.
This imprecision in handling the data leads to a gradual decrease in
coherence of the signal measured at two elements with increase in $\Delta \nu$, 
the distance of the source from the phase center, and the baseline length.
For a point source, the effect of fractional bandwidth and baseline length on 
the reduction in peak response ($I/I_0$, where $I_0$ is the peak response) can 
be conveniently parameterised in terms of a dimension parameter $\beta$ defined 
as $\Delta \nu/\nu_0\!\times\!\theta_0/\theta_{HPBW}$, where $\theta_0$ is the 
distance of the point source from the phase center measured in units of the
half power beam widths for a given baseline ($\lambda_0/D$, where $\lambda_0$ 
is the wavelength corresponding to $\nu_0$ and $D$ is the length of the baseline) 
\citep{1999sira.conf..371B}.
Substituting for $\theta_0$, $\beta$ can be expressed as $\Delta \nu\ D\ \theta_0/c$.
When $\beta = 1$ the peak response decreases to $\sim\!0.8$ and further reduces 
to $\sim\!0.5$ when $\beta=2$.
A necessary requirement for synthesis imaging is that all the measured visibilities
used to reconstruct the sky brightness distribution receive coherent emission 
from the same physical patch of the sky.

Typically, interferometer baselines span a wide range while the bandwidths over
which individual visibilities are measured is a constant.
This leads to visibilities from different baselines receiving correlated emission 
from sky patches centered at a common spot but differing in size.
At high radio frequencies this poses no problems because the FoV gets limited by the
diffraction beam of the aperture before the effects of the coherence in the
radiation received at different elements reduced significantly.
On the other hand, at VLF frequencies, as discussed in the preceding sections, 
the FoV of the receptors is necessarily very large. 
A reasonable solution to the problem of matching the patches from where correlated
emission is received is to measure visibilities over sufficiently narrow bandwidths 
so that even the longest baseline receives correlated flux from the entire FoV.

At high frequencies, for most part the background emission is so weak that sky can 
be regarded as {\em cold} and the number density of radio sources is such that the 
sky can be considered to be largely {\em empty}. 
Hence, the common practice of not imaging the entire FoV but only a small parts of 
it from where the emission is expected is not only acceptable but also prudent.
At VLF frequencies, the sky background is extremely intense.
This implies that in order to obtain the best image fidelity and dynamic range 
performance, it would be necessary to image the entire region from which correlated 
flux is received, which, in the present case corresponds to the entire FoV. 

Mapping a large primary beam also requires caution to be exercised on
another front. 
The geometric delay, $\tau_g$, suffered by signal arriving from 
different directions changes widely, from $0\ sec$ perpendicular to 
baseline to $D/c\ sec$ along the baseline.
The process of cross correlation, however, allows for correction of 
only a unique geometric delay, $\tau_g^{\prime}$. 
If the residual geometric delay, $\tau_g - \tau_g^{\prime}$, exceeds 
$1/\nu_{ch}$ for some directions, where $1/\nu_{ch}$ is the coherence 
time of a band limited signal of bandwidth $\nu_{ch}$, the signal 
received from these directions will loose correlation. 
This effectively limits the field of view which can be mapped by the 
interferometer. 
Such a situation can be avoided by correlating the signal over 
sufficiently narrow channel widths so that the inequality mentioned 
above is never satisfied. 
As an example, a baseline length of a $100\ km$ requires that the
channel width be $\le 3.0\ kHz$.

\subsection{Telemetry considerations}
\label{subsec:telemetry1} 
The telemetry bandwidth available to the VLF interferometer to
transmit the data to the Earth is an important design driver.
We estimate the largest telemetry bandwidths expected to be available 
to a VLF interferometer in near future to lie in the range $5$ to 
$10\ Mbps$ (Mega bits per second).
This is based largely on the fact that the ALFA study 
\citep{2000GMS...119..339J} expected to sustained a $8\ Mbps$ telemetry 
bandwidth from an array of spacecraft located $10^6\ km$ from the 
Earth, using the existing $11m$ subnet of NASA Deep Space Network 
(DSN) facility.

All the earlier proposals for space
based VLF interferometers chose to transmit the Nyquist sampled
time series from every receiver on each of the satellites to the
Earth. This classical approach offers the advantages of simpler
satellite architectures and a homogeneous array which provides
ample redundancy. The principal disadvantage however is that
because of the very voluminous nature of the data, these designs 
can only provide a rather limited observation bandwidths. 
The ALFA proposal, for instance, provided a maximum of $125\ kHz$, 
using a $8\ Mbps$ telemetry down-link and a $1\ bit$ quantisation 
for the data stream \citep{2000GMS...119..339J}.

In order to preserve the expected dynamic range of about $70\ dB$ 
in the input signal \citep{1996AdSpR..18...35B}, a $12$ bit sampling 
is required.
This large number of bits required per sample further reduce the 
available bandwidth of observation by close to an order of magnitude.
There is no doubt that a larger bandwidth of observation than what
the classical approach can provide in near future would be very 
desirable. 
There are only two ways in which this can be achieved, either by 
simply waiting for the telemetry technology to progress sufficiently 
to meet the needs of VLF interferometry or by doing some on-board
data processing to reduce the data volumes.
For an interferometer, the data products which can be meaningfully 
averaged, are the visibilities. 
For most ground based instruments the process of computing 
visibilities is performed by dedicated custom made hardware 
of significant complexity.
The hardware capabilities required for on-board computation of 
visibilities have remained significantly beyond the state of the art 
till recently. 
Therefore, in spite of the low bandwidth disadvantage,
transmitting the Nyquist sampled time series had been the only
feasible option for the earlier VLF space interferometer designs.

Judging from the current trends in the industry, the computational
capabilities of space qualified hardware are expected to increase
at a rate considerably larger than at which the telemetry
bandwidths are expected to grow. 
We, therefore, consider it judicious to assess the option of 
on-board visibility computation and subsequent time and frequency 
averaging to reduce the data volume and increase the available
bandwidth of observation.


\subsection{Propagation effects}
\label{subsec:propagation-effects}
At the VLF frequencies, the inhomogeneous, turbulent and magnetised 
plasma of ISM and interplanetary medium (IPM) 
act like mediums with refractive index fluctuating in both space 
and time.
The propagation of the VLF radiation from distant radio objects 
through this medium modifies the incident wavefronts in a considerable 
manner.
The implications of most relevant of these propagation effects 
are enumerated.
\begin{enumerate}
\item {\em Angular broadening} -- The passage of VLF radiation through ISM 
and IPM results in significant apparent angular broadening of compact 
sources \citep{2000rickett}. 
This angular broadening effectively imposes a limit on the finest
resolution with which one can expect to study the universe,
irrespective of the baseline lengths involved. 
The angular
broadening scales as the square of the wavelength of the radiation
and is hence most severe at the VLF range.
Fig. \ref{fig:resolution} shows the angular broadening of a point
source due to IPM and ISM as function of frequency.
The resolution afforded by baselines of $50$ -- $100\ km$ is also 
shown to provide a point of reference.
\begin{figure}
\vspace{-0.8cm}
\begin{center}
\includegraphics[scale=0.77, angle=0]{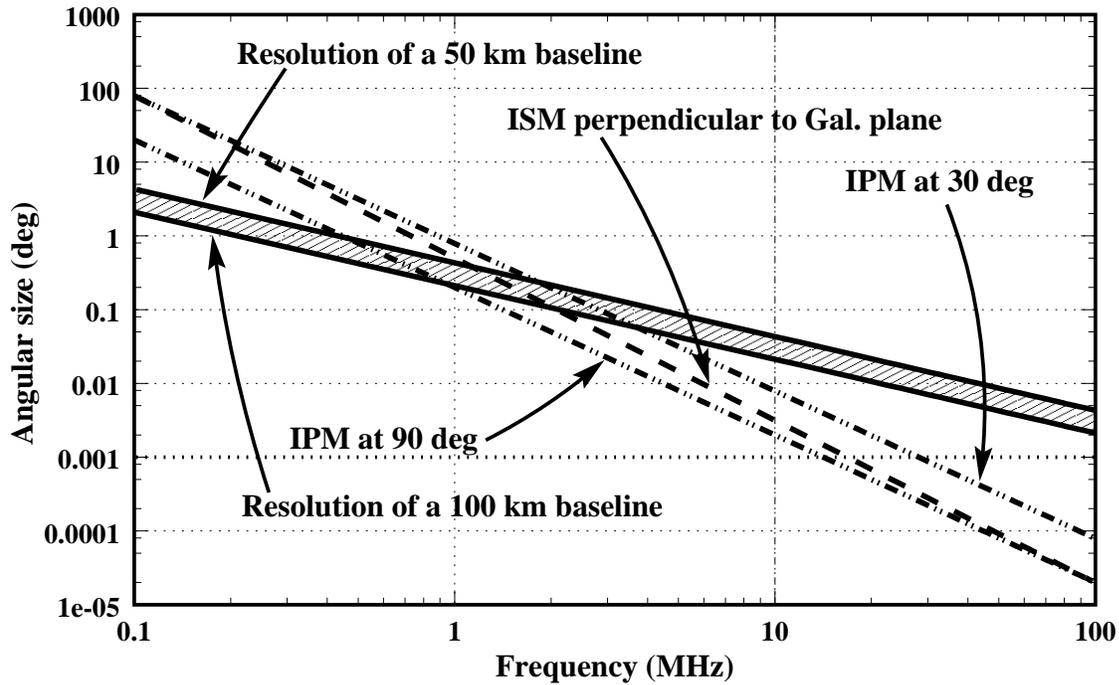}
\caption{{\bf Angular broadening due to ISM and IPM} -- The dashed
line shows the angular broadening of a point source due to the ISM
in a direction perpendicular to the Galactic Plane. The effect in
the Galactic place can be two to three orders of magnitude larger.
The dashed-dotted lines show the angular broadening of a point
source due to the IPM at elongations of $30^{\circ}$ and $90^{\circ}$ 
respectively. 
The shaded region is bounded by the resolution of baselines of 
length $50$ and $100\ km$. }
\label{fig:resolution}
\end{center}
\end{figure}

\item {\em Temporal broadening} -- The scattered rays which lead to
angular broadening of compact sources travel through different
paths to the observer. 
This leads to a spread in the travel time of signal, which
results in smearing of transient signals like pulses from a
pulsar.
Due to the comparatively much larger distances involved, interstellar
temporal broadening is much more severe ($\sim\! 5\ yr$ at $1\ MHz$) 
than interplanetary ($0.1\ sec$ at $1\ MHz$) \citep{2000woan}.

\item {\em Depolarisation of radiation} -- The magnetised nature of the 
ISM and the IPM, their large inhomogeneities combined with the fact that 
magnitude of Faraday rotation is proportional to $\lambda^{2}$ conspire 
to make this an important effect for VLF range.
The multi-path propagation from source to observer, with possibly 
a different Faraday rotation for each path, presents a fundamental 
limitation to the detection of linearly polarised signal. 
Unlike the case at higher frequencies, this cannot be treated
like simple foreground Faraday rotation which can be estimated
by multi-frequency measurements.
Circular polariastion, however, is not effected by Faraday rotation
though the intrinsic luminosity of cyclotron processes which 
produce it is much lower than that of synchrotron processes.
A detailed discussion on polarisation effects can be found in
\citet{1996AJ....111.2465L}.

\item {\em Absorption effects} -- The free-free absorption is expected
to render the ionised ISM optically thick at some turnover 
frequency.
This frequency will be a function of both, the emission measure of the
medium and its electron temperature.
The Warm Ionized Medium is expected to turn optically thick for 
path lengths of about $2\ kpc$ at $3\ MHz$.
As the Galactic disc is about $1\ kpc$ in thickness, the sky
will appear to be foggy in all directions at frequencies of a 
few MHz.
It will be possible to see out of the Galactic plane at higher
frequencies and the apparent appearance of the plane itself will be 
dominated by the mottling due to the presence of discreet regions
of high electron density.
The subject is discussed in considerable detail by \citet{2000ralw.conf..257D}.

\item {\em Reflection, refraction and scattering close to the Sun} -- 
Due to the large gradient in the electron density in high Solar corona, a 
variety of unusual reflection and refraction phenomenon take place 
in this region.
An increasingly large fraction of the Solar corona becomes inaccessible 
to the radio waves at frequencies below $\sim\!20\ MHz$ as ray paths get
reflected back into the IPM. 
\cite{1956ApJ...123...14B} give an in depth discussion of this 
and a few other interesting phenomenon.
Close to the Sun, there is considerable evidence, from multi-spacecraft 
studies of solar bursts below $1\ MHz$, for the existence of anomalous 
beaming and large angular scattering \citep{1989A&A...217..237L}.
\end{enumerate}

We note that of the effects mentioned here, only scatter broadening
effects impact the design of a VLF interferometer, all the others limit 
the science which can be done at these frequencies independent of the 
design details.

\subsection{The radio frequency environment}
\label{subsec:rfi1} 
A look at the frequency allocations chart for the United States of 
America shows that the {\em entire} VLF band from $\sim\! 9\ kHz$ onwards 
is allocated to specific users, with the exception of $13.36$ -- $13.41\ MHz$ 
and $25.55$ -- $25.67\ MHz$ which are reserved for radio astronomy.
The spectral allocation situation is expected to be similar in other 
parts of the world.
\citet{1990lfas.work...59E} concluded in his study that the radio frequency
interference (RFI) in near the near Earth environment was too strong to allow 
sensitive VLF interferometric observations.
Measurements from the WIND satellite, located $\sim\!2\!\times\!10^5\ km$ 
from the Earth, in the frequency range $1$-$15\ MHz$ reveal that 
in spite of the geometric dilution and the attenuation offered by the 
ionosphere, RFI is often $5\ dB$ stronger than the Galactic background 
emission \citep{1996GeoRL..23.1287K}. 
A strong correlation between the spectral bands allocated to commercial 
shortwave stations and the RFI affected parts of the spectrum was found.
However the $20\ kHz$ wide spectral channels of on-board instrumentation did 
not provide sufficient resolution to routinely identify individual broadcast 
services, which are a few $kHz$ wide and spaced about $5\ kHz$ apart. 
It was also evident that commercial shortwave transmission does not 
account for all the observed RFI.
Given the grossly insufficient coverage provided by the protected bands,
a VLF radio interferometer will inevitably have to operate in a rather 
unfriendly RFI environment.
It will hence be necessary to develop a RFI mitigating strategy.

\subsection{Non-stationarity of the sky}
\label{subsec:non-stationary-sky}
Often, the instantaneous sampling of the {\em u-v} plane achieved by an
interferometric array falls short of the requirements for a good image of
the desired part of the sky and does not provide sufficient sensitivity.
Ground based instruments rely on rotation of the Earth to improve the 
{\em u-v} coverage and observing for longer durations improves the 
sensitivity.
The space borne VLF interferometer will similarly rely on the motion 
along its orbit and changes in baselines due to relative velocities 
between the constellation elements to improve its sampling of the {\em u-v-w} 
volume and the sensitivity.
Using visibilities collected over a period of time to construct a
single map of the sky implicitly assumes time stationarity of the
sky over the period of observation. 
On the time scales of operation of the VLF interferometer mission, 
the evolution of most astronomical sources is a non issue and we may 
choose to ignore the low frequency variability, weather intrinsic or 
introduced by the propagation effects. 
However, the apparent positions of the solar system objects, with 
respect to the more distant objects, change rapidly.
The position of the Sun, for instance, changes by about $1^{\circ}$ 
per day.  
This implies that the sky sampled by the interferometer at different 
epochs corresponds to different realisations of the sky, differing in 
the location of the solar system objects with respect to the more distant 
ones. 
The active Sun, Earth and Jupiter are among the stronger discreet 
sources in the VLF range and their emissions have an elaborate 
frequency-time structure.
In addition, the apparent angular sizes of sources will vary in 
time, as their angular distance from the Sun changes (Sec. 
\ref{subsec:propagation-effects}, Fig. \ref{fig:resolution}).


\section{The design concept}
\label{sec:strawman}
In order to keep the mission economically feasible a {\em
micro-satellite} based approach has been chosen so that all the
interferometer elements can be deployed using a single launch
vehicle. The interferometer is envisaged to comprise of a
constellation of about $16$ free floating three axis
stabilised micro-satellites. Each member of the
constellation will serve as an interferometer element. As
motivated in section \ref{subsec:telemetry1} we investigate
the possibility of an on-board correlator. This has the
consequence that not all the spacecraft will be identical. One of
them will receive the data streams from all the others and will perform
on-board Digital Signal Processing (DSP) to reduce the data
volumes to be transmitted to the Earth. We refer to this
spacecraft as the {\em Mother} spacecraft. In order to avoid
a single point of failure in the design, it will be necessary 
to equip a few of the satellites, say three, to take up the 
role of the Mother spacecraft. 
Some details of a design based on this concept follow along with 
brief justifications of the choices made.

\subsection{Frequency coverage}
\label{subsec:freq-coverage}
The mission has been designed to cover the frequency range [$0$ --
$40\ MHz$]. The frequency range, useful for radio astronomy, is limited on
the lower end by the plasma frequency of the IPM, known to be about a 
few tens of $khz$ at $1\ AU$. 
Close to the plasma frequency, interferometric measurements are so badly
corrupted by the propagation effects that they no longer remain
useful for studying distant radio sources. 
It may be possible to conduct meaningful interferometric observations
down to $\sim\!0.1\ MHz$. At the high frequency end, we
expect that by about $30\ MHz$, it will be scientifically more 
rewarding and much more economical to use the more powerful and
versatile upcoming ground based low frequency instruments like
LOFAR. None the less, we strongly advocate an overlap in the 
frequency ranges covered by the space array and the ground based
instruments and suggest $40\ MHz$ as the upper frequency limit for
the space array (Fig. \ref{fig:LFinst}). 
The overlap in frequency range will help in
calibration and allow the space array to benefit from the
information of the sky obtained by the ground based array. 
A reliable and detailed model for the low frequency sky obtained by
the ground based instruments will provide a very good anchor point
from where to boot strap to proceed to lower frequencies.
We note that the instrument itself will be designed to work below
the plasma frequency of the IPM and the measurements in this part
of the spectrum can serve as local plasma measurements which can 
be used to serve different scientific objectives.
Their discussion lies beyond the scope of this paper.

\subsection{Receiving elements}
\label{subsec:receptors}
As mentioned in section \ref{subsec:largeFoV}, for a VLF interferometer,
the long wavelengths involved, the constraints of a space borne mission 
and a micro-satellite based architecture limit the choice of elements to 
short dipole antennas.
Due to their poor noise characteristics, short dipoles are usually not the
preferred choice for receiving elements in radio astronomy.
However, the intense Galactic background emission in the VLF range
and the enormously wide primary beam of the short dipole ensure
that the the noise on the measured signal is dominated by that due to 
the Galactic background emission and not the receiver noise
\citep{2000ralw.conf..329M}.
The dipoles could be based on the monopole stacer design used for the 
WAVES instrument on board the WIND satellite \citep{1995SSRv...71..231B} 
or the ones designed for SWAVES on board STEREO.
These antennas are $6\ m$ in length and the noise contribution of the 
antenna itself is less than that from the Galactic background in a 
frequency range from $\sim\!400\ kHz$ to $\sim\!40\ MHz$ providing a
good match to the needs of a VLF interferometer.
The choice of the interferometer element specifies the FoV or the 
primary beam size for the interferometer and for a short dipole it 
is $\sim\! 8\pi/3\ sr$, or about $27.5\!\times\!10^3\ deg^2$.
Each satellite will be equipped with three mutually orthogonal short
dipoles, in order to record all the information in the
electromagnetic field incident on the satellite. 

The use of three mutually orthogonal dipoles offers some advantages 
over the conventional use of two mutually orthogonal ones -- computing 
all the nine ($3\!\times\!3$) cross-correlations per baseline allows 
one to construct Stokes parameters to characterise the polarisation of
radiation received from any arbitrary direction 
\citep{2000PhRvE..61.2024C}, as opposed to being limited to directions 
close to the perpendicular to plane defined by the two dipoles; on being 
equipped with the additional ability to compute all the nine 
auto-correlations, the use of three orthogonal dipoles permits individual 
satellites to be used for direction finding of polarised sources 
\citep{1995RaSc...30.1699L}, a potentially useful feature for initial 
deployment of the constellation and for calibration; and lastly the use 
of independent data from a third dipole can be considered as an 
increase in the effective collecting area or an effective reduction in 
observation time needed to achieve a given sensitivity.

\subsection{The signal path and on-board Digital Signal Processing}
\label{subsec:signal-path+DSP}
The signal from each of the three short dipoles on every
constellation elements is fed via a low noise amplifier to an analog 
to digital converter (ADC). 
The ADC Nyquist samples the signal at $80\ MHz$ to cover the 
entire radio frequency (RF) range of interest. 
The input signal must be sampled with sufficient bit depth to
to preserve its fidelity.
\citet{1996AdSpR..18...35B} suggested that, for a short dipole, the 
dynamic range of the input signal is expected to be $60$ to $70\ dB$.
We aim for a $70\ dB$ dynamic range, which requires sampling
using $12$ effective bits.

The primary guiding principle for on-board Digital Signal Processing 
(DSP) approach is to distribute it to the largest extent possible,
in order to avoid a build up of DSP requirements at some later stage 
in the signal chain. 
The three digitised time series on each of the spacecrafts will be 
Fourier transformed in real time.
The spectral width of the frequency channels is determined by the 
length of the longest baseline and the requirement of imaging the
entire primary beam (Sec. \ref{subsec:map-full-FoV}).
A maximum baseline of $\sim\!100\ km$ (Sec. \ref{subsec:propagation-effects}) 
and a requirement that the reduction in peak response due to decorrelation 
loss be less than $20\%$ ($\beta \lesssim 1$; Sec.~\ref{subsec:map-full-FoV})
for near $4\pi\ sr$ fields of view, lead to a bandwidth of $\sim\!1\ kHz$ 
for the width of the spectral channels.

The DSP required to achieve this spectral resolution will be implemented 
as a two stage fast Fourier transform (FFT) engine.
As an illustration, the first stage takes a $512$ point real transform 
and yields and $256$ point complex spectra with a spectral resolution
of $156\ kHz$. 
A second stage will perform a 128 point complex Fourier transform on 
a subset of these spectral channels leading to $1.22\ kHz$ wide 
spectral channels.
Performing the second stage FFT on $25\%$ of the spectral channels 
delivered by the first stage of FFT leads to a requirement of 
$\sim\!1\!\times\!10^9 CMACS$ ({\em complex multiplications and additions
per second})
per polarisation per satellite, for a $80\ MHz$ sampling.
The number of channels on which the second stage FFT will be performed
will depend on computing power available on-board and the intra constellation
telemetry bandwidth limitations.
A selected subset of these narrow spectral channels will be re-sampled 
using one or two bits and transmitted to the Mother spacecraft.

The Mother satellite will receive the Nyquist sampled spectral data 
from all the constellation members and will compute the auto and cross 
correlations.
The resulting visibilities will be averaged over suitable intervals 
in frequency and time.
Say $25\%$ of the available $1.22\ kHz$ spectral channels are resampled
using one or two bits and transmitted to the Mother satellite. 
This leads to a requirement of $\sim\!3\!\times\!10^9\ CMACS$ for the
computation of a set of 9 ($3\!\times\!3$) cross and self correlations
for each baseline.
Keeping in mind that the operations on the individual constellation
members are done on $12\ bit$ data and most of those on the Mother 
satellite are done on $1$-$2\ bit$ data, the total on-board computing 
requirements for the Mother satellite will be about twice as much as
those for other constellation members for the $2.5\ MHz$ RF bandwidth
provided by the above design.
The correlator itself will be a flexible and reconfigurable device.
It will allow a range of combinations of spectral and temporal resolutions
and the number of baselines for which the correlations are computed, while
keeping the total throughput from the correlator a constant and  respecting
the constraint of available telemetry bandwidth to the Earth.
For instance, it will be possible to get higher temporal and/or spectral 
resolution at the cost of decreasing the bandwidth of observation and/or 
number of baselines used.
It will be desirable for the correlator to have the ability to respond 
to self generated and external triggers in order to switch to an 
appropriate temporal and spectral resolution mode in response to an event.
The averaged visibilities will finally be transmitted to the Earth,
where the rest of the analysis will take place.
The temporal extent over which the visibilities can be averaged will be
decided by the shorter of the time scales at which it is desired to study
the received emission and the relative velocities of the constellation 
members which lead to a gradual change in baselines.
The spectral averaging extent will depend on the requirement for spectral
resolution and the intrinsic spectral characteristics of emission.

At first sight, the tasks of digitising a 40 MHz wide band and
Fourier transforming it into $kHz$ wide spectral channels seems like
an unlikely task for a micro-satellite. 
The present day technology, however, comes very close to meeting 
these requirements.
According to its data sheet, the best performing space qualified 
$12\ bit$ ADC from Analog Devices available in July 2003, AD9042, 
can sustain a maximum sampling rate of $41\ MHz$, has a typical power 
dissipation of $595\ mW$ and provides a spurious free dynamic 
range of $80\ dB$ over $20\ MHz$.
It seems likely that by the time it is required, the available technology 
will allow the signal to be oversampled in order to recover some of 
the losses in digitisation process.
Space qualified FPGAs with $6\!\times\!10^6$ gates are already available 
from Xilinx and their road map for the near term future promises
$1.5\ V$ Virtex-II Pro family devices with $> 10^7$ gates by the end of 
2004.
For the example configuration discussed here, just one of these devices 
per spacecraft will comfortably be able to handle the DSP requirements 
for all three dipoles.
Preliminary studies indicate that it may even be possible to accommodate
the additional computing requirements of the Mother spacecraft on the 
same device or may at most require the use of another similar device.
Meeting the on-board DSP requirements of a VLF interferometer in near
future does not seem to pose a significant problem.

\subsection{Telemetry and bandwidth of observation}
\label{subsec:telemetry+bw} 
The telemetry issues can be sub divided into those relating to 
intra constellation telemetry and telemetry from the Mother spacecraft
to the Earth.
We first discuss the former.
As mentioned in section \ref{subsec:telemetry1}, the limitation on 
the telemetry bandwidth to the Earth led us to consider the approach 
of reducing data volume on-board.
However, it is impossible to reduce the data rates to below Nyquist
requirements before the computation of visibilities.
As a consequence, all constellation members transmit Nyquist rate data 
streams to the Mother spacecraft.
It is instructive to compute this number per $MHz$ of RF bandwidth.
For each constellation member transmitting data to the Mother spacecraft,
this amounts to $2\! \times 10^6 \times N_{dipoles} \times n_{bits} 
= 6\ Mbps/MHz$, with a $N_{dipoles}$, the number of dipoles per spacecraft,
of 3 and $n_{bits}$, the number of bits per sample, of 1.
For a constellation of $N_{craft}$ spacecrafts, the Mother spacecraft 
receives data from $N_{craft}-1$ spacecrafts simultaneously. 
For the 16 element constellation under consideration, this implies a rate 
of $90\ Mbps/MHz$.
The telemetry rate grows by an order of magnitude to accommodate the 
$10\ MHz$ RF bandwidth which the on-board DSP can comfortably deliver.
This design, therefore, has intra constellation telemetry requirements in
the range of $0.1$ -- $1\ Gbps$.

The telemetry requirements for transmission of data from the Mother
satellite to the Earth depend on the spectral and temporal averaging 
performed on-board.
The amount of data to be transmitted to the Earth per second is given 
by the following expression
\begin{equation}
T = \frac{N_{craft}(N_{craft} - 1)}{2} \times N_{corr} \times N_{bits} \times 
\frac{\Delta \nu_{RF}}{\nu_{ch}} \times
\frac{1}{\tau_{int}} \times \frac{1}{N_{ch}}
\end{equation}
where the first term gives the number of baselines for a constellation
comprising of $N_{craft}$ spacecrafts, 
$N_{corr}$ is the number of correlations computed for each baseline, 
$N_{bits}$ is the number of bits used to represent each complex visibility, 
$\Delta \nu_{RF}$ is the RF bandwidth for which visibilities are computed, 
$\nu_{ch}$ is the width of each of the spectral channels, 
$\tau_{int}$ is the integration time and $N_{ch}$ the number of 
spectral channels which are averaged over.
For a $\Delta \nu_{RF}$ of $1\ MHz$, using a $N_{craft}$ of 16, $N_{corr}$ of 9, 
$N_{bits}$ of 16, $\nu_{ch}$ of $1.22\ kHz$ and averaging over $10$ 
spectral channels for $10$ seconds leads to a data rate of $142.8\ kbps$.
For a $10\ MHz$ RF bandwidth, it grows to a modest $1.43\ Mbps$ and the 
entire RF bandwidth accessible to the instrument, $40\ MHz$, requires only 
$5.71\ Mbps$.
The vast reduction is data volume achieved by on-board data processing is
apparent on comparing the rates at which flow in and out of the the Mother 
spacecraft or by recalling that the ALFA design required $8\ Mbps$ of 
telemetry for a RF bandwidth of $125\ kHz$.
The huge reduction of telemetry requirements to the Earth offers another 
benefit as well.
The fact that the data rate from earlier mission designs were was essentially 
limited by available telemetry implied that the observations could be made 
only for the duration of the telemetry down-link, requiring a 24 hour down-link
for continuous observations.
With the large reduction in the telemetry requirements, it might now become
possible to reduce the down-link duty cycle without compromising the observing 
duration.

The RF bandwidth over which the visibilities are finally computed will be 
determined by the most limiting bottle neck in the data path.
In view of the large intra constellation telemetry requirements of this design, 
we expect it to be the most limiting resource.
In order to make the most judicious use of this scarce resource, the $12\ bit$ 
spectra available at individual spacecrafts are re-sampled using $1\ bit$ before 
transmission to the Mother satellite, as mentioned in section 
\ref{subsec:signal-path+DSP}.

The final sensitivity achieved by the design will depend on the bandwidth 
available for intra constellation telemetry.
The design provides a theoretical point source sensitivity of $5.6\ Jy$ and 
$2.0\ Jy$ at $3$ and $30\ MHz$, respectively, for $1\ MHz$ of bandwidth and 
$1\ min$ time integration.

It can be argued that this design simply moves the telemetry 
bottleneck from constellation-Earth telemetry segment to the 
intra constellation telemetry segment.
It is true that the data volume cannot be reduced below Nyquist 
requirements before the computation of visibilities. 
The only way to reduce the data rate is by reducing the RF bandwidth 
covered.
We believe that the primary reason for this bottle neck to exist is that
this functionality was never needed till now and not that it is
intrinsically a difficult problem to solve.
The space industry is now enthusiastically considering formation flying 
missions.
As more missions involving multiple spacecrafts needing to communicate
with another and exchange information in real time come up, this requirement 
will get addressed and suitable technological solutions will emerge.

\subsection{Constellation configuration}
\label{subsec:configuration}
The spatial configuration of the constellation needs to be tuned to 
the needs of VLF interferometry (Sec. \ref{sec:vlf-interferometry}).
It must take into account the nature of the VLF sky, the scientific 
objectives and the engineering constraints.
For instance -- the angular resolution of the constellation will be 
limited in most directions by the angular broadening due to IPM and 
ISM beyond baselines of about $80$ -- $100\ km$ 
(Fig. \ref{fig:resolution}),
in order to be sensitive to the intense large angular scale solar and 
galactic background emission, short baselines ranging from a few $\lambda$ to 
a fraction of $\lambda$ are needed and
the near isotropic beam and the requirement to simultaneously map
the entire FoV require a very good $u$-$v$-$w$ coverage. 
It is not entirely clear if it is more advantageous to aim for a
complete {\em and} uniform coverage or compromise on completeness in favour
of a Gaussian fall off in the $u$-$v$-$w$ coverage density.
In this paper, we have chosen to aim for a complete and uniform 
$u$-$v$-$w$ coverage. A promising configuration for achieving this 
was presented in the ALFA mission proposal \citep{2000GMS...119..339J}.
The spacecraft were distributed on a spherical surface in a pseudo random 
manner while respecting a minimum separation constraint between the nearest 
neighbours.
The $u$-$v$-$w$ coverage provided by such a configuration, referred to 
as an {\em Unwin sphere} in the literature, is remarkably uniform and 
isotropic in nature, and can provide good $4\pi$ synthesis imaging 
capabilities.

Ideally, the resolution provided by a VLF interferometer should be close to
the limit set by scatter broadening in all directions.
However, the angular broadening due to scattering by the IPM is a strong function of 
the angular distance from the Sun for elongations less than $90^{\circ}$. 
Close to the Sun, the angular broadening due to the IPM is so large that
baselines larger than $\sim\!5$--$10\ km$ are expected to completely
resolve out the solar emission.
While in the directions far from the Sun, even a 100 km baseline might not be 
scatter broadening limited.
This introduces a strong anisotropy in the VLF synthesis imaging resolution 
requirements.

Given the limited number of instantaneous baselines, for a mission with
significant solar science objectives, it would be desirable to have all
of them to be sensitive to solar emission.
For an Unwin sphere kind of constellation this suggests a maximum diameter 
of $\sim\!8\ km$ for the sphere, leading to a resolution of $43'$ at $3\ MHz$ 
\footnote{A baseline of $17.2\ km$ provides a resolution of $\sim\!1^{\circ}$ at $1\ MHz$.}.
In view of the anisotropy in the angular resolution requirements, it is worthwhile
to consider the possibility of incorporating a corresponding anisotropy in the 
Unwin sphere configuration.
In order to be able to provide higher angular resolution in other directions
while keeping all the projected baselines, as seen from the Sun, sufficiently
small, we propose to distribute the satellites in a quasi random manner on a 
{\em Cigar} shaped surface, respecting a similar minimum separation constraint
as the Unwin sphere.
The cigar can be about $\sim\!80\ km$ in length and about $\sim\!8\ km$ 
across at it's center and oriented such that it always points in the 
direction of the Sun. 
Rather than regarding the spacecraft being placed on a geometric 
surface, it is more appropriate to regard them as being placed within a 
cigar shaped shell of finite thickness.
Over a period of time, as the constellation goes around the Sun on it's 
orbit, the long baselines which lie along the length of the cigar will 
sweep through a range of orientations, providing a good coverage of the
$u$-$v$-$w$ volume needed for high resolution imaging of the celestial 
sphere.

An obvious limitation of this configuration is that all the long baselines lie 
in the ecliptic plane.
Assuming that the total number of spacecraft in the constellation is a 
constant, the only way to remedy this is by removing some of
the spacecraft from the cigar and deploying them such that they provide
long baselines perpendicular to the ecliptic plane.
Removing even a few spacecraft from the cigar reduces the number of 
useful baselines for solar observations considerably.
Given the limited number of interferometer elements available, it is 
difficult to meet all these demands simultaneously.

An alternative approach for resolving the conflicting requirements of 
solar and astronomical imaging is that rather than trying to find less
than perfect, compromise solutions for both sets of objectives simultaneously, 
divide the mission duration between the two and try to meet the requirements 
of only one of them at a time.
An Unwin sphere whose radius a slowly increases from say $\sim\!5\ km$ to
$\sim\!80\ km$ over the mission duration is a good solution for this 
approach. 
At small diameters the constellation is capable of fulfilling solar 
objectives and as the diameter grows, its ability to do solar science
diminishes.
The data gathered using the small diameter phase will, of course, be
useful for astrophysical objectives as well, if a satisfactory solution to
the problem of non-stationarity of the sky can be found 
(\ref{subsec:non-stationary-sky}).
The gradual increase in the constellation radius will also provide  
a better $u$-$v$-$w$ coverage.

A detailed study, examining the imaging characteristics of different 
configurations and their compatibility with different scientific 
objectives, is needed to arrive at a suitable constellation configuration 
for a VLF interferometer.
As has been mentioned earlier, the width of the spectral channel is
related to the maximum baseline in the configuration.
The current choice of $\sim\!1\ kHz$ wide spectral channels allows
for maximum baseline lengths of $\sim\!100\ km$ beyond which 
decorrelation losses may be considered significant (Sec. 
\ref{subsec:telemetry+bw}).
If the final configurations are much smaller in size, the spectral
widths of the frequency channels can be increased leading to an 
increased RF bandwidth coverage while using the same telemetry 
bandwidth between the Mother spacecraft and the Earth.

\begin{figure}
\begin{flushleft}
\includegraphics*[bb = 50 200 515 375, scale = 1.11, angle=0]{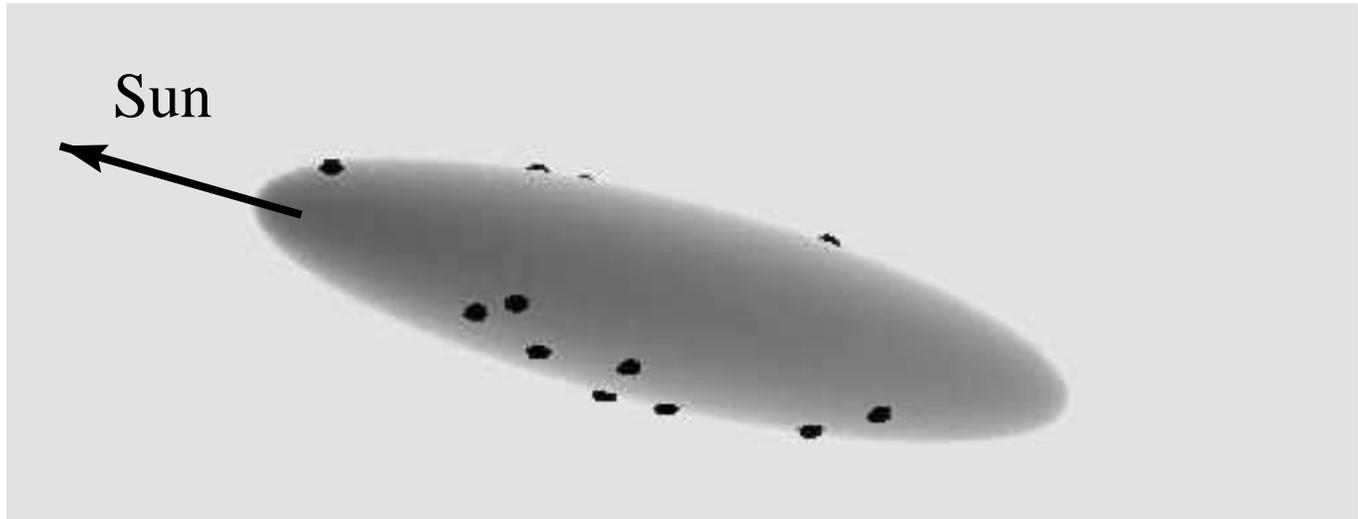}
\caption{{\bf The constellation configuration} --
A possible Cigar shaped constellation configuration is shown.
The spacecraft are shown as dark spots on the Cigar shaped surface.
}
\label{fig:cigar}
\end{flushleft}
\end{figure}

\subsection{Tackling the Radio Frequency Interference}
\label{subsec:rfi2}

As discussed in section \ref{subsec:rfi1}, RFI will be a dominant 
issue for near Earth orbits and will remain an issue to contend with
even for the far Earth orbits.
There are two aspects of RFI seen by space based instruments, which 
make the problem different in character from the one faced by ground 
based instruments.
\begin{enumerate}
\item A space based instrument benefits from the $1/r^2$ geometric 
dilution of the intensity of the RFI emission, where $r$ is the 
distance between the Earth and the satellite borne interferometer.
\item Unlike being immersed in a {\em sea} of RFI, as on Earth, a 
space based instrument sees RFI to be coming from a specific direction
in the sky, that of the Earth. 
To provide a point of reference, from a distance of 
$\sim\!1.0\!\times\!10^6\ km$, the Earth will remain unresolved by 
$\sim\!25\ km$ baselines at $1\ MHz$. 
\end{enumerate}
It is also important to keep in mind that in spite of practically 
the entire VLF band being allocated to specific users, the RFI 
spectral occupancy is much less than $100\%$ (probably closer to 
$50\%$) because of the frequency separation between transmissions 
on adjacent allocated frequency channels.
The RFI ridden VLF band is interspersed by regions of relatively 
clean spectrum which can be utilised for radio astronomy. 
The $kHz$ wide spectral channels provide a good match to the frequency 
resolution needed to make use of the parts of the band between 
adjacent transmissions.
We proposed the following to mitigate the harmful effects of RFI.
\begin{enumerate}
\item In order to maximise the geometric dilution of RFI, it would
be preferable to choose from among all available orbits the ones 
which place the constellation farthest from the Earth.

\item The well defined directional nature of RFI will be exploited
to identify and reject parts of the band with strong RFI.
The astronomical observing will be periodically interrupted, at a 
low duty cycle, for RFI detection. 
The interferometer will sweep through the RF band of interest and 
will be configured to add the signal coming from the direction of 
Earth first constructively (in phase) and then destructively (out 
of phase). 
RFI is expected to show up as narrow band emission which appears 
when the signal from Earth is added in phase and drops when it is 
added out of phase.
The differences in the spectra obtained in the two cases will be 
examined to identify RFI contaminated frequency channels, which 
will then be excluded from further processing.
Frequency and time integration can be performed to improve the
sensitivity of RFI detection, though at the price of a reduction
in the time available for astronomical observations. 
By averaging in time for a minute and over four adjacent frequency 
channels, we expect to obtain $3\sigma$ detection of RFI at the 
level of $20 \%$ of the Galactic background, which is quite 
satisfactory. 
The frequency with which this exercise is performed should match
the time scale at which the RFI environment is expected to change.
The data from WAVES experiment on-board WIND can be useful in 
this determination \citep{1996GeoRL..23.1287K}.
As this scheme needs to examine the visibilities to identify RFI,
it needs to be implemented on the Mother spacecraft.
For optimal utilisation of the intra constellation telemetry 
bandwidth resource, a list of spectral channels identified as RFI 
contaminated will be transmitted to the constellation members.
These identified channels will not be transmitted to the mother
spacecraft and the list will be updated every time RFI detection
is performed. 

\item While it is possible to identify the relatively strong RFI 
in short time integrations, it is much tougher to identify weaker 
RFI and prevent it from contaminating the data. 
Fortunately for a space borne array, its highly directional nature 
ensures that the residual RFI which sneaks in will map on to a 
specific predictable direction in the sky, the direction of the Earth. 
This direction will trace out the path followed by the Earth in 
the sky, as seen from the array. 
Hence rather than contaminating the entire map, the RFI signal 
will stay confined to this locus of the Earth through the sky.
As the effect of RFI is localised in the image domain, it suggests 
that it will be useful to an entirely new class of RFI mitigation techniques, which work in the image domain, can be 
usefully employed.
\end{enumerate}

\subsection{Choice of orbits}
\label{subsec:orbit}
In order to keep it feasible to maintain the constellation in the
desired configuration, for the entire duration of the mission, it
is necessary to restrict the choice of orbits to those where the
differential gravity over the length scales of the constellation
size is low. 
These orbits also allow the visibility data to be averaged for 
significant durations in time (many tens of seconds), reducing 
telemetry bandwidth requirements.
This becomes possible because the relative positions of the 
spacecrafts, and hence the baselines, evolve only slowly in time.  
Distant Earth orbits provide comparatively better RFI environments 
as well (Sec. \ref{subsec:rfi2}). 
The above considerations argue strongly in favour of distant Earth 
orbits.
On the other hand, distant Earth orbits pose a tougher telemetry
problem.
None the less, we consider only distant Earth orbits to be suitable 
for this mission, especially in view of the fact that the proposed 
design reduces the telemetry intensive nature of the mission in a 
very considerable manner. 
Halo orbits about the L1 Lagrange point, distant retrograde and 
prograde orbits about the Earth-Moon barycenter at distances of 
$\sim\! 10^6\ km$ from the Earth seem to be suitable candidates.
The choice of orbits in the vicinity of the L1 Lagrange point 
imply that the interferometer will always be facing the sunlit side 
of the Earth. 
This will lead to a better shielding from the extremely intense 
terrestrial auroral kilometric radiation \citep{1979JGR....84.6501G} 
and will be advantageous for the solar and astronomical scientific 
objectives.


\section{Data analysis strategy}
\label{sec:data-analysis} 
The primary focus of this work is to present a VLF interferometer 
design optimised to provide visibilities with as little loss of 
information as possible and maximises the RF bandwidth of 
observation. 
The intent is to make the task of imaging more tractable and better 
address the scientific objectives.
The inversion of the visibilities received on the Earth to arrive at 
the brightness distribution in the sky will be a non-trivial task
and forms an independent area of research.
The problem of inverting VLF interferometer visibilities is 
more involved than the one faced by existing interferometers because
of the a near $4\pi\ sr$ FoV (Sec. \ref{subsec:map-full-FoV}) and 
the fact that the non-stationarity of the sky will need to be taken 
into account (Sec. \ref{subsec:non-stationary-sky}).

The earlier proposals for VLF interferometers sought, without much
success, to reduce the FoV to be mapped by trying to use more 
directional elements \citep{1988A&A...195..372W,1997RaSc...32..combined}.
As it is not possible to build directional receptors smaller 
or comparable in dimensions to the wavelength of operation, 
the space based VLF interferometer will have no other option but
to have practically omni directional beams, till the time technology 
allows the deployment of structures, a few $km$ in size, in space.
A later attempt intended to use the bandwidth decorrelation effect
to limit the region of sky from which correlated radiation is received
\citep{2000GMS...119..339J}.
This is not an entirely satisfactory solution to the problem because,
as was pointed out in Sec. \ref{subsec:map-full-FoV}, this leads 
visibilities coming from different baselines to pick up correlated
emission from sky patches of different orientations and sizes.

We wish to point out that these attempts to limit the fields
of view to be mapped were driven by the fact that it was inconceivable 
then to envisage that the computing power required to image near all-sky
fields of view will become available in the foreseeable future. 
Hence, in spite of being mathematically well formulated, the task of 
simultaneous all sky imaging was never feasible. 
The recent enormous increase in the computational power available at 
affordable costs and the continued adherence to {\em Moore's Law} 
will soon bring the computational requirements for all-sky mapping with 
reach, eliminating the need for reducing the region to be mapped. 
It is now time to start considering the simultaneous all sky coverage 
provided by the near omni-directional beams as an advantage over 
other wavelength ranges in astronomy which are limited to narrower
fields of view.

We propose to map the entire primary beam of receiver elements.
The visibilities obtained will be gridded over the three
dimensional {\em u-v-w} space and will lead to a three dimensional
{\em image volume} using the full three dimensional inversion
formalism \citep{1999sira.conf..383P}. 
The intersection of the image volume with a unit sphere represents 
the celestial sphere, the object of interest. 
The dimensions of the visibility cube on which to grid the data will 
be arrived at using the criteria that the resulting image volume should 
Nyquist sample the entire sky with the resolution of the constellation. 
The choice of mapping the entire primary beam impacts the design of 
the interferometer in a considerable manner. 


The data analysis techniques must have the ability to deal with the 
{\em non-stationarity} of the sky.
As the Sun will be the strongest non-stationary source, the bulk of 
the problem can be diminished by simply resolving out solar emission. 
This is however not compatible with solar science objectives and 
the small baselines are required to capture the intense galactic 
background.
Strictly speaking the data accquired by the VLF array over a period 
of time cannot be combined to produce a single sky map without first
removing from it the contamination due to the non-stationary sources.
One possible way of doing it might be to use an all-sky model for the 
stationary sky and subtract it's contribution from the measured
visibilities. 
The residual visibilities will then represent the contribution of the
non-stationary part of the sky, which is also expected to have
significant time and frequency structure.
The overlap in frequency range with ground based instruments will 
provide the VLF interferometer a good sky-model from where to 
boot-strap to lower frequencies.


\section{Calibration and formation flying }
\label{subsec:calibration}
The aim of any calibration procedure is to estimate the response 
function of the measuring instrument. 
In synthesis imaging this has conventionally been done by observing 
astronomical sources with known properties (position, strength, 
structure, polarisation etc.).
This approach relies on the availability of a field of view which is
dominated by a single (or at most a few) strong sources with known
properties. 
The near $4 \pi\ sr$ field of view of the present instrument coupled 
with the intense Galactic background in the VLF range eliminate this
possibility, except in case of some solar bursts, which usually have
an elaborate frequency time structure. 
It will, hence, not be possible to routinely rely on astronomical 
sources for calibrating the interferometer. 
There will be no choice but to seek engineering solutions to meet 
the calibration requirements.

Complex gain of the receivers will be calibrated by periodically 
injecting a known complex calibration signal into the signal path 
just after the dipoles and comparing the output from the receiver 
with the input signal.
Comparisons with observations from the ground based instruments at 
the higher end of the VLF range and simultaneous observations of
bright transients like solar bursts by other space and ground based
instruments can be used for amplitude calibration.
\cite{2001A&A...365..294D} have shown that for the low resolution
single element instruments at low frequencies, the Galactic background
spectrum offers a more reliable means for flux scale calibration.
This can be used to obtain an independent estimate the gains of 
individual constellation members.
The complex primary beam of the dipoles will have to be measured 
on the ground, before the launch of the mission. 
Mounting the dipoles on the spacecraft will considerably modify
the beam shapes from those of isolated dipoles.
The beam shapes must, hence, be characterised after the dipoles have 
been mounted on spacecraft.
In flight calibration of beam shapes, if needed, will require some 
additional on-board functionality.
It can be achieved by radiating a signal of known characteristics 
from one or few of the constellation members, receiving it on the 
others and processing it suitably.

The formation flying requirements relate directly to the wavelength
of operation of the interferometer.
Working at the longest possible wavelengths, a VLF interferometer
has the least demanding formation flying requirements.
This makes a VLF interferometer an ideal choice for testing the 
formation flying mission control and management concepts.
While infra-red space interferometers, like DARWIN 
\citep{1996Ap&SS.241..135L} require the constellation members to be 
positioned with relative accuracies of the order of a $cm$, for a
VLF interferometer it is not necessary to fly the constellation 
in a rigid pre-defined configuration at all. 
Departures of individual spacecrafts from their intended positions,
by small fractions of the characteristic length scale of the configuration, 
do not degrade the performance on the interferometer in any significant 
manner.
The mission requires the baselines to be calibrated to an accuracy
of $\sim\!0.1\ \lambda$ at the smallest wavelength of observation ($0.75\ 
cm$ at $40\ MHz$).
It is however not necessary to know the relative positions of the 
constellation members to this accuracy in real time. 
Real time baseline accuracies only need to be sufficient to ensure 
that the correlation between the signal received at different 
satellites is not reduced significantly.
Real time accuracies of the order of a few $m$ will be quite acceptable. 
The final accuracies, made available by the {\em post facto} 
orbit reconstruction, must however meet the more stringent 
requirement. 
Once the baseline are known to final required accuracies, offline
corrections can be applied to the computed visibilities.
Being far away from the Earth, it will not be possible for the
mission to make use of the existing Global Positioning System
(GPS) satellite network for locating the satellites. 
An on-board ranging and direction finding system will be required.

The intra constellation ranging and direction finding data is 
insensitive to an overall rotation of the array, which will need
to be determined by independent means. 
There is also a need to align the relative orientations of 
constellation members (attitude control), to ensure that the
dipoles on different spacecrafts are pointed in the same directions.
A star tracker unit on board every constellation member will be 
used to serve both these purposes.
As the FoV of dipoles are huge, attitude control of the order of 
a degree is probably already an over kill.

The time synchronisation requirements for the constellation do not
pose a challenge.
As the design does not require a local oscillator, there is no 
requirement to distribute a phase locked signal to all spacecrafts.
The only requirement for time synchronisation comes from the coherence
time of the signal, which is $\propto 1/\nu_{ch}$.
For channel widths of $\sim\! 1\ kHz$, this amounts to an embarrassingly 
comfortable $ms$.
Given that the relative positions of the spacecrafts will be known 
to a few $m$, it will be preferable to time tag the time series of 
spectra from different constellation members on the Mother spacecraft, 
taking into account the changing light travel time and the fixed delay 
due to the signal processing.



\section{Conclusions}
\label{sec:conclusions}
Very low frequency radio astronomy is now coming of age. 
The advances in technology have finally brought us at the brink 
of opening this last unexplored window in the electromagnetic
spectrum, both from ground and space.
A versatile and powerful low frequency interferometer, LOFAR, is 
envisaged to commence operation later this decade, pushing low 
frequency radio astronomy to its furthest limits achievable from 
the ground.
A space based array will benefit considerably from the knowledge of 
the low frequency sky gained from this ground based array and will 
form the next logical step of pushing the exploration of the VLF
window to its absolute limits.
Exploiting the advances in space qualified technology and the vast
increase in computing capabilities, one can now design space borne 
VLF interferometers with capabilities way beyond the earlier proposals. 
The currently available technology comes very close to meeting the
requirements of this new design in most respects.
We believe that the returns from this approach will amply justify 
the rather short wait for the technology to deliver the needs of a
VLF interferometer.
This design offers many advantages over the conventional approach.
It permits the RF bandwidth covered to be increased by more than an
order of magnitude.
The simultaneous digitisation of the entire RF bandwidth of interest 
offers the flexibility of distributing the part of it to be processed 
further in any manner from a few $kHz$ to $40\ MHz$,
a considerable advantage for multi frequency synthesis and for 
simultaneously covering a large spectral window.
A flexible and reconfigurable correlator design can be used to set 
observation parameters to suit different science needs and evolve 
the observing strategy as we learn during mission operation.
The data it provides is compatible with the scientific objectives of
the mission.

A disadvantage of the design is that as it requires a Mother spacecraft, 
the redundancy in the constellation is considerably reduced.
To counter this, it will be essential to equip a few spacecraft to take 
up the role of a Mother spacecraft in case of need.
This does not seem to pose an unsurmountable problem.
This design does require a more complex payload than the conventional 
approach, but the level of complexity is not large in an absolute sense.
The intra constellation telemetry is presently envisaged to be the most 
constraining bottle neck.
This is an area which has received little attention by the space industry 
till now, simply because none of the existing space missions needed this
functionality.
Formation flying missions are now being vigourously pursuing by the space
industry and the requirement of intra-constellation telemetry will be 
shared by many of them.
As this requirement gets better recognised we expect suitable technological 
solutions to emerge.

The steady increase in performance of space qualified hardware,
formation flying and computing capabilities, place the demands of
a satellite based interferometer within reach in near future. It
is, hence, timely to lay down a design for a space low frequency 
interferometer, based on  realistic expectations for technology 
available in near future and assess its scientific desirability. 
The VLF interferometer concept is now mature enough to merit a 
detailed engineering study.
\\


\begin{acknowledgements}
We acknowledge the fruitful and illuminating discussions with 
Alain Kerderaon, Claude Mercier, Sanjay Bhatnagar, Pramesh Rao, 
Tim Bastian, Brian Corey, Will Aldrich and Brian Fanous at various 
stages during this study. 
This research has made use of NASA's Astrophysics Data System 
Abstract Service. 
All the this work was done using the GNU/Linux operating system 
and it is a pleasure to thank the numerous contributors to this 
software.
\end{acknowledgements}


\bibliography{/home/doberoi/LPCE/NETRA/bib/netra}

\end{document}